\documentclass[reprint]{revtex4-1}

\usepackage{bbold} % For unit matrix 1, \mathbb{1}, to work
\usepackage{graphicx}% Include figure files
\usepackage{mathtools}
\usepackage{url}
\usepackage{amsthm} % For proof
\usepackage[final]{changes} % [final]
\usepackage{hyperref} % For hyperlinks
\usepackage{xfrac}
\usepackage{nicefrac}

\begin{document}

\title{Depletion force between disordered linear macromolecules}

\author{Nathaniel Rupprecht, Dervis Can Vural}
\affiliation{University of Notre Dame, South Bend, IN}

\date{\today}

\begin{abstract}
When two macromolecules come very near in a fluid, the surrounding molecules, having finite volume, are less likely to get in between. This leads to a pressure difference manifesting as an entropic attraction, called depletion force. Here we calculate the density profile of liquid molecules surrounding a disordered \added{rigid} macromolecules \added{modelled as a random arrangement of hard spheres on a linear backbone}. We analytically determine the position dependence of the depletion force between two such disordered molecules by calculating the free energy of the system. We then use molecular dynamics simulations \added{to obtain the depletion force between stiff disordered polymers as well as flexible ones and compare the two against each other}. We also show how the disorder averaging can be handled starting from the inhomogenous RISM equations.
\end{abstract}

\maketitle

Objects immersed or dissolved in a liquid will experience an emergent attractive force, as the liquid molecules, having finite volume, cannot squeeze between them \cite{likos2001effective}. Put another way, it is entropically favorable for the objects to be close, since each have surrounding volumes unavailable to the liquid molecules; and when objects approach to the extent that these volumes overlap, the molecules have more volume to explore \cite{asakura1954interaction,asakura1958interaction}. Thus objects are more likely to be near each other, as if they attract. This entropic force is called ``depletion.''

Since the seminal papers of Asakura and Oosawa, depletion forces has had far reaching implications from molecular physics \cite{minton1981excluded,gholami2006entropic,mravlak2008depletion,maghrebi2011entropic,sapir2015depletion} and biochemistry \cite{hall2003macromolecular,hanlumyuang2014revisiting,braun2016entropic}, to high energy physics \cite{visser2011conservative,basilakos2014entropic,typel2016variations,feng2016effects,fadafan2016entropic}. It has even been suggested that gravity \cite{verlinde2011origin,plastino2018entropic,plastino2018quantum,bhattacharya2018comments} and the Coulomb force \cite{wang2010coulomb} might be entropic forces. So far, depletion forces between plates immersed in rods \cite{mao1997theory,chen2002depletion} or spherocylinders \cite{mao1997density}, forces between colloids \cite{crocker1999entropic}, semiflexible chains \cite{castelnovo2004semiflexible}, spherocylinders \cite{li2005depletion}, and ellipsoids \cite{konig2006depletion,miao2014depletion} immersed in colloids, and forces between colloids immersed in polymer \cite{striolo2004depletion}  have been established.  Forces mediated by mixtures of two types of particles have also been studied \cite{triplett2010entropic}. 

Depletion interactions are particularly relevant in polymer physics \cite{fuchs2001macromolecular,fuchs2002structure,tuinier2015depletion,likos2016depletion,perez2017colloid}. \added{The} Ornstein-Zernike \added{(OZ)} equation \cite{ornstein1914integral,scholl2003self,brader2013nonequilibrium}, Reference Interaction Site Model (RISM) \cite{misin2015communication,tormey2016rism,johnson2016small}, and Polymer RISM (PRISM) \cite{schweizer1994prism,schweizer1993analytic,perry2015prism} yields correlation functions between flexible polymers, rods, and colloids. Generally, additional assumptions are needed to close \added{OZ and RISM-type} equations, since they each contain multiple functions whose forms are not specified by the defining equation  \cite{baxter1968percus,yethiraj1992self,martynov1999new,saumon2012quantum}\added{. Closure equations specify properties of the functions involved in PRISM, and include the Percus-Yevick approximation \cite{percus1958analysis} and the hypernetted-chain equation \cite{fries1985solution}. These \added{models and closure relations} are then} typically solved numerically, though some analytical methods also exist \cite{wertheim1963exact,wertheim1964analytic,yasutomi2000analytical,adda2008solution,rohrmann2011exact}.

\begin{figure}
    \centering
    \includegraphics[width=0.47\textwidth]{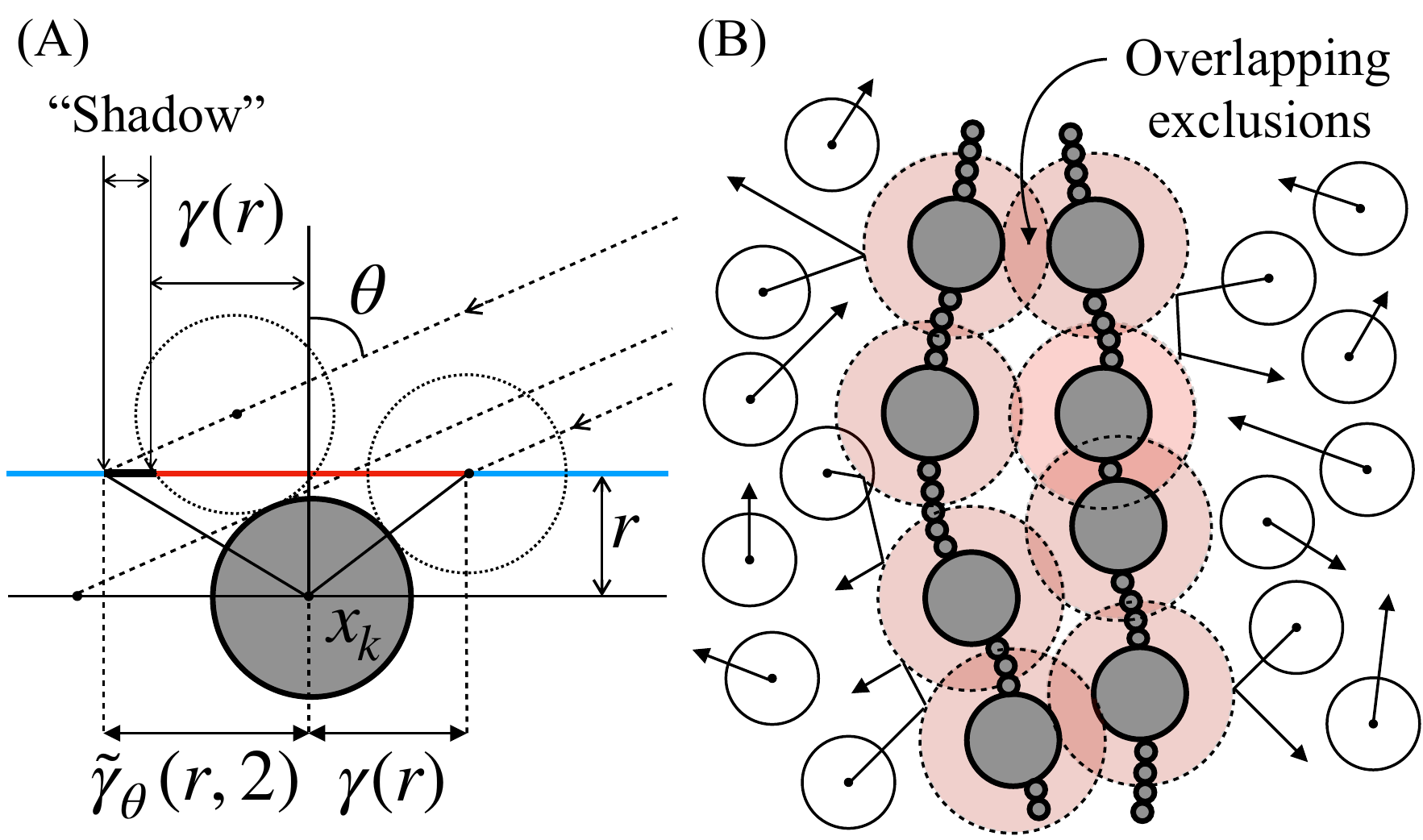}
    \caption{\textbf{(A):} If an incident sphere manages to approach \(r\) above the backbone, it will not be in the excluded regions. A distance \(\gamma(r)\) to the right and \(\tilde{\gamma}_\theta(r)\) to the left of each sphere is excluded in this case.
    \textbf{(B):} \added{A depiction of flexible polymers and how depletion forces arise. The proximity of polymers increases the chances that the large monomers overlap, which is entropically favorable since it allows the hard sphere fluid to occupy more space.}
    }
    \label{Fig:setup}
\end{figure}

\added{Depletion forces are also important in biophysics, where they play a crucial role in the organization of cells, particularly in chromosomal organization and compaction \cite{stavans2006dna,jeon2016molecular,jeon2016effects,pereira2017entropic,shendruk2015simulating,kumar2019impact}, the behavior of DNA inside cells \cite{odijk1998osmotic}, crowding-induced chromatin compaction \cite{oh2018entropic}, and polymer chain looping \cite{bian2019unusual}.
}

While depletion forces always originate from disordered arrangements of a solvent, the objects experiencing the force themselves have always been chosen by authors to be orderly geometric shapes, such as planes, cylinders, squares, and spheres. In polymer physics, while RISM allows for the incorporation of intramolecular structure, the structure function can only represent specific objects rather than an ensemble of disordered objects.

Here, we study the depletion force between two \emph{disordered} objects. Specifically, we consider rigid \added{and flexible} disordered polymers (a random arrangement of hard spheres) immersed in a solvent (also hard spheres). 
These objects can be seen as a model of disordered block copolymer, where one monomer is very small and the other is large \cite{stepanow1996copolymer,westfahl2005correlated,kim2018origins}.
We first find the probability that a sphere incident towards a \added{straight} polymer will approach by \(r\) before any collision. This is the disorder averaged polymer-liquid correlation function in the low density limit. We then use this to evaluate the average depletion force between two \added{straight} polymer chains. \added{Molecular dynamics simulations are used to verify our theoretical results. We then numerically calculate the depletion force for the more general case of flexible polymer chains.} Finally, we show how RISM can be extended to treat \added{the straight disordered polymers}.

\textbf{Liquid profile near a disordered chain.} We model a rigid disordered polymer as a line of length \(L\) with \(N\) non-overlapping, but otherwise randomly-placed spheres of radius \(R_1\), and the surrounding liquid molecules as spheres of radius $R_2$.
\added{
Consider a liquid molecule colliding with one of the monomers as shown in Fig. \ref{Fig:setup}A, with a dotted empty circle on the right, and a filled gray circle respectively. We define our coordinate system such that the polymer lies on the $x$-axes horizontally, and that the monomer and liquid molecule collide at a point where the vertical coordinate of the liquid molecule is \(y = r\). At the point of contact, the difference in the \(x\) coordinates between the monomer and the liquid molecule is
}
\begin{align}
    \gamma (r, \kappa) = \begin{cases} 
      \kappa \sqrt{1 - \left(r/\kappa\right)^2} & r \le \kappa \\
      0 & r > \kappa
   \end{cases}.
    \label{StoppingFunction}
\end{align}
where \(\kappa = 1+R_2/R_1\). \added{The ``stopping function,'' $\gamma (r, \kappa)$, is useful because it tells us that a liquid molecule \(r\) away from the backbone cannot have a horizontal distance less than \(\gamma(r,\kappa)\) to a monomer. This region is excluded since the two disks would have to overlap.}

\added{Here and throughout \(\gamma(r)\) and \(r\) are defined in units of \(R_1\). For the rest of the paper we pick $R_1=R_2$ for cosmetic and pedagogical reasons, however the $R_1\neq R_2$ case is a straightforward generalization. Also} note that much of our analysis can be further generalized to non-spherical shapes by replacing \(\gamma\) with another appropriate stopping function. 

\added{We also define a second related quantity \(\tilde{\gamma}_\theta(r,\kappa)\). We consider a liquid molecule incident at angle $\theta$ that passes by the monomer tangentially, as shown in Fig. \ref{Fig:setup}A with a dotted empty circle on the left. When $r$ away from the backbone, its horizontal distance to the monomer is defined as \(\tilde{\gamma}_\theta\). Note that \(\tilde{\gamma}_\theta \ge \gamma\) due to the ``shadow'' of the disk on the line. We will evaluate \(\tilde{\gamma}_\theta\) in a moment.
%At certain angles, the disk can also be launched so that it is able to reach a height of \(r\), but does so by first coming into tangential contact with the disk on the line. The launched disk does not collide with disk on the line, but if it were any closer, it would. In Fig. \ref{Fig:setup} A, the dotted unfilled disk on the left will not be in contact with the gray disk when it reaches a height of \(r\), since to reach this height without colliding with the gray disk, it must pass above the gray disk. The x distance between the the point where the disk would be at a height \(r\) and the center of the disk on the line is defined to be \(\tilde{\gamma}_\theta(r,\kappa)\) and is also an excluded region. Note that this depends on the angle at which the disk was launched, and that \(\tilde{\gamma}_\theta \ge \gamma\) due to the ``shadow'' of the disk on the line. We will evaluate \(\tilde{\gamma}_\theta\) in a moment.
}

To estimate the liquid density profile around the chain, we calculate the probability that a molecule incident towards the chain at an angle \(\theta\) can approach to a distance \(r\) without contact. Given a configuration of spheres at coordinates \(\{x_0,...,x_{N-1}\}\), and a incidence angle \(\theta\), we will count up the fraction of the interval \([0, L]\) in which the incident sphere could start to make it to a  distance at least \(r\) above the central axis (cf. Fig \ref{Fig:setup}). We will then integrate over all valid configurations of spheres. 

The measure of configurations of spheres on a line is equivalent to the configurations of hard lines on a line, known as a Tonks gas \cite{tonks1936complete,giaquinta2008entropy}. We write it as
\begin{align}
\Gamma_N &= \!\!\int\! d\mathbf{x} \!\equiv\! \frac{L}{N} \prod_{k=1}^{N-1} \!\int_{x_{k-1} + 2 \added{R_1}}^{L-2 \added{R_1} (N-k)}\hspace{-0.3in} dx_k = \frac{L^N}{N!}(1-\phi)^{N-1}
\label{GammaIntegralForm}
\end{align}
\added{where \(\phi \equiv 2 N R_1 / L\) is the (linear) volume fraction of the disks on the line, i.e. the amount of the line that is covered by disks.}

As the incident sphere approaches the chain, it will at some point be at a distance \(r\) to the backbone. The \(x\) coordinate of the incoming sphere when it is \(r\) above the backbone will be labeled \(x_p\). We will approach the computation by conditioning on the interval that \(x_p\) falls into.

\begin{figure}
    \centering
    \includegraphics[width=0.47\textwidth]{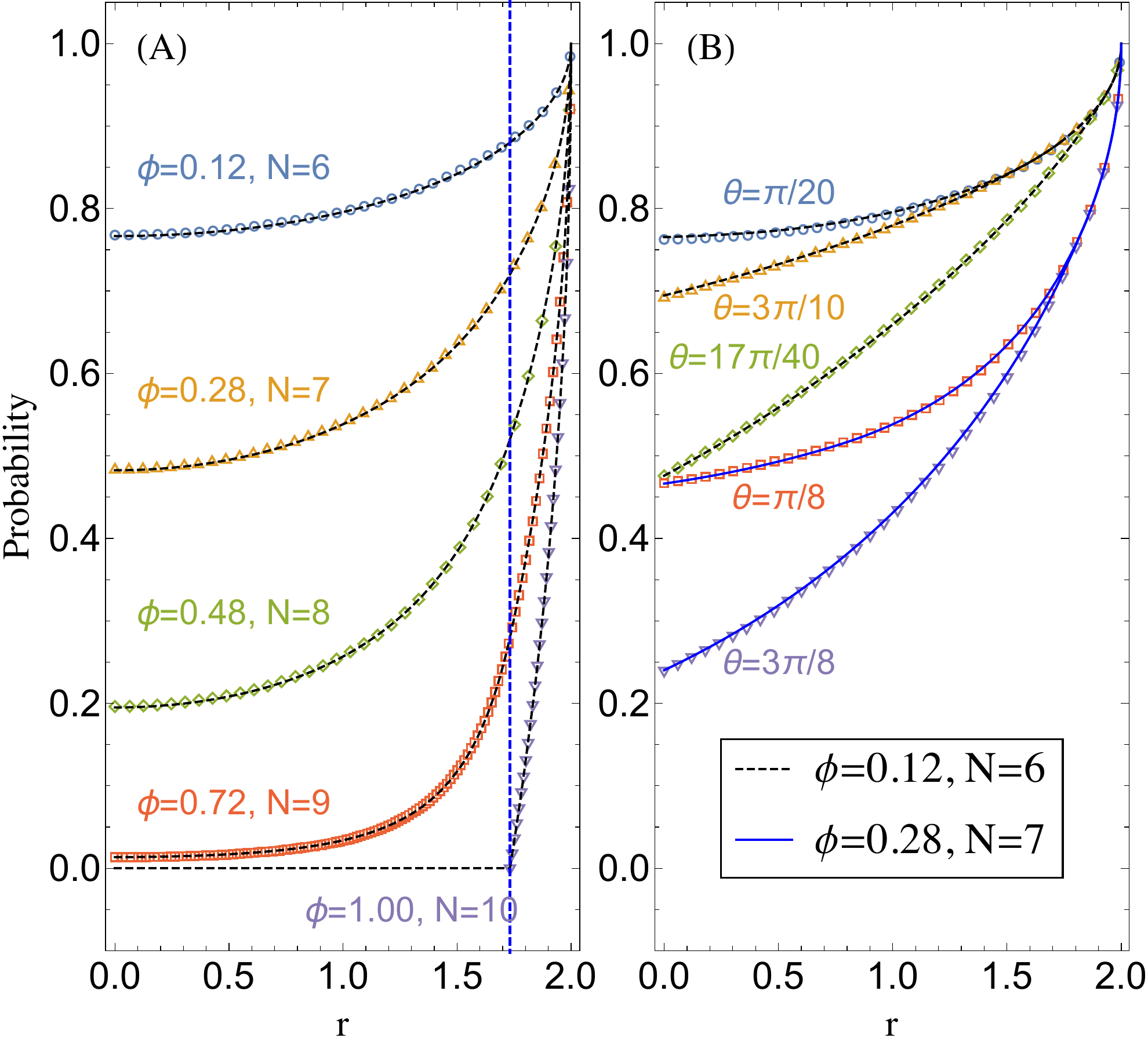}
    \caption{Comparison of analytical result eqn.(\ref{ApproachProbabilityFinite}) (red dashes) with simulations (dots) where spheres incident towards a chain at \(\theta=0\) (A) and various angles (B) for $L=1$. The vertical dashed line represents the point where the piece-wise function (\ref{ApproachProbabilityFinite}) transitions at \(\gamma=1\), for \(r=\sqrt{3}\).
    \label{Fig:Numerics}}
\end{figure}

Each sphere in the chain excludes a distance \(\added{R_1}\gamma(r)\) to its right, and \(\added{R_1}\tilde{\gamma}_\theta(r,\kappa)\) to its left, where 
\begin{align}
\tilde{\gamma}_\theta(r, \kappa) = \begin{cases}
0 & r \ge \kappa 
\\
\gamma(r, \kappa) & \kappa > r \ge r_0
\\
(r_0 - r) \tan \theta + \gamma(r_0, \kappa) & r_0 > r \ge 0
\end{cases} 
\end{align}
where \(r_0 = r_0(\theta) \equiv \kappa   \sin \theta\). The case for \(\kappa \tan \theta >r \) emerges because of the ``shadow'' the excluded volume that a sphere casts on the line.

Let \(x_p \in [x_k, x_{k+1}]\) \added{(note that these \(x\)'s are have dimensionality of length)}. Then probability that it is able to approach to a distance \(r\) above the line is proportional to the length \([x_k, x_{k+1}]\) that is not excluded by \(\gamma\) or \(\tilde{\gamma}\)
\begin{align}
Pr(r \vert x_p \in [x_k,  x_{k+1}] )
= \frac{\Xi \big(x_k - x_{k+1} - 2 \added{R_1} \xi(r) \big)}{x_{k+1}-x_k}
\end{align}
where \(\xi(r) \equiv [\gamma(r) + \tilde{\gamma}(r)]/2\), and the clamp function \(\Xi(x)\) is \(\Xi(x)=x\) for \(x>0\) and 0 otherwise. The clamp function is necessary, because the excluded areas of \(\gamma\) and \(\tilde{\gamma}\) may overlap.
Since the particle's intersection position, \(x_p\) will be in \([x_0, x_1]\) \emph{or} in \([x_1, x_2]\), etc., and the particle is initialized uniformly at random,
\begin{align}
    Pr(r   \vert \vec{x}) &= \sum_{k=0}^{N-1} Pr(r \vert x_p \in [x_k, x_{k+1}]) Pr(x_p \in [x_k, x_{k+1}])
\nonumber \\
    & = \frac{1}{L}\sum_{k=1}^N \Xi \big(x_k - x_{k-1} - 2 \added{R_1}\xi(r) \big)
\end{align}
Therefore, to obtain \(p_\phi(r, N) = Pr(r)\) we just need to integrate over all the arrangements of \(x_k\),
\begin{align}
\!p_\phi(r, N) &= \frac{1}{\Gamma_N} \int d\mathbf{x} \frac{1}{L} \sum_{k=1}^N \Xi \left(x_k - x_{k-1} - 2 \added{R_1}\xi(r) \right)
\label{PassageProbabilityIntegralForm}
\end{align}
where the integration measure is the same as (\ref{GammaIntegralForm}), and \(1/\Gamma_N\) comes from the normalization of the measure, since we are averaging over all valid configurations of spheres.

If \(r>2\), then \(p_\phi(r, N) = 1\). If \(r\) is large enough that \(\xi(r) \le 1\), then every \(\Xi\) function is non-zero, and the sum in (\ref{PassageProbabilityIntegralForm}) telescopes and can be evaluated easily,
\[
\frac{1}{L} \sum_{k=1}^N \Xi \big(x_k - x_{k-1} - 2 \added{R_1}  \xi(r) \big) = 1 - \phi \, \xi(r)
\]
which is independent of any of the positions of the spheres on the line. This term then pulls out of the integral, which cancels with the factor of \(1/\Gamma_N\), leaving us with 
\begin{align}
p_\phi(r, N, \theta) = 1 - \phi \, \xi(r) \mbox{ for } \xi(r) \le 1.
\label{ApproachProbabilityFar}
\end{align}
If \(\xi(r) > 1\), integrating (\ref{PassageProbabilityIntegralForm}) is more involved, since the clamp functions can be zero (see \ref{Appendix:CalcDetails}),
\begin{align}
p_\phi(&r,N,\theta,\kappa) =  \label{ApproachProbabilityFinite}\\
& \begin{cases}
1 - \phi\,\xi(r) & \xi(r) \le 1 
\\
\left(1 - \phi \right) \left( 1 - \phi \frac{\xi(r)-1}{N (1-\phi)} \right)^N & 1 < \xi(r) \le (\frac{N}{\phi}-N)+1\
\\
0 & N(\frac{1}{\phi}-1)+1 < \xi(r)
\end{cases}\nonumber
\end{align}
In the thermodynamic limit \(N,L\to\infty\),  \(N/L = \phi/(2R)\),
\begin{align}
& p_\phi(r, \theta, \kappa) = \begin{cases}
1 - \phi \, \xi(r) & \xi(r) \le 1 \\
(1-\phi) \exp \left[ \frac{\phi \left(1-\xi(r) \right)}{1-\phi} \right] & \xi(r)>1
\end{cases} 
\label{ApproachProbability}
\end{align}
Note that \(\xi(r)\) is actually a function of \(\kappa\), \(\theta\), and \(r\)\added{, and therefore dimensionless}. 

\added{To confirm this result, we carry out 2D simulations where we launch 50,000 disks from random positions towards a line of randomly arranged disks, and recording the \(r\) value at which they hit. We do this at both \(\theta=0\) (Fig. \ref{Fig:Numerics} A) and at various nonzero \(\theta\)'s (Fig. \ref{Fig:Numerics} B) and find excellent agreement between analytic formulas and simulations.}

The approach probability (\ref{ApproachProbability}) can be viewed in several additional ways. One way of looking at (\ref{ApproachProbability}) is that its limiting case is the probability that there is a gap of size \(x = \xi(r) /2\) in a Tonks gas. Indeed, the probability that there is a gap of size \(x\) at a generic point in a Tonks gas in the thermodynamic limit is \cite{torquato1990nearest,elkoshi1985one,giaquinta2008entropy}
\[
P(x) = (1-\phi) \exp [\phi(1-2x)/(1-\phi)].
\]

\begin{figure}
    \centering
    \includegraphics[width=0.49\textwidth]{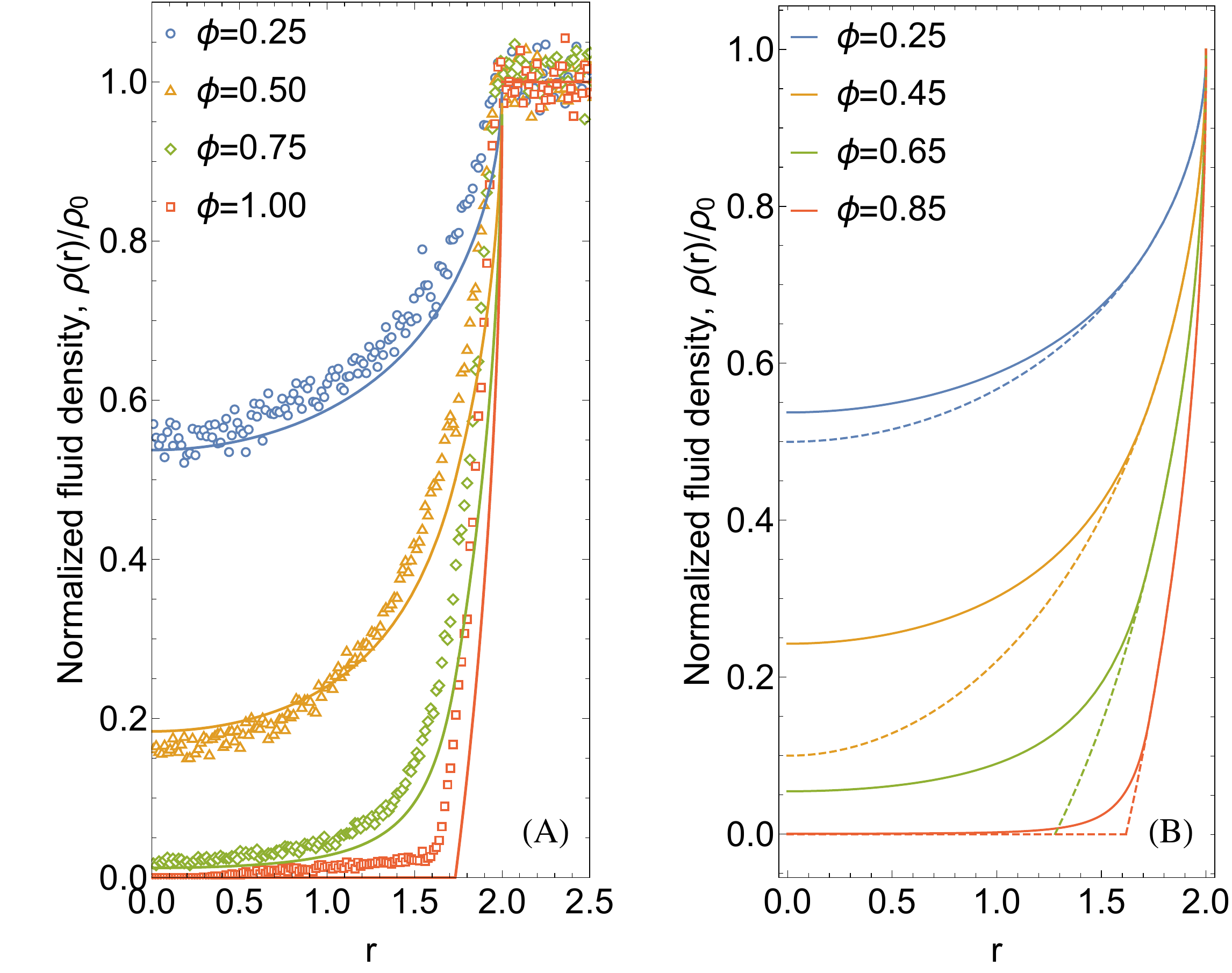}
    \caption{(A) Comparison of theory (lines) and simulation (dots) for a rod-like polymer. (B) Comparison of disordered (lines) and ordered (dashes) polymers. In both (A) and (B), interactions between fluid molecules are neglected.}
    \label{Fig:Polymer}
\end{figure}

\begin{figure}
    \centering
    \includegraphics[width=0.49\textwidth]{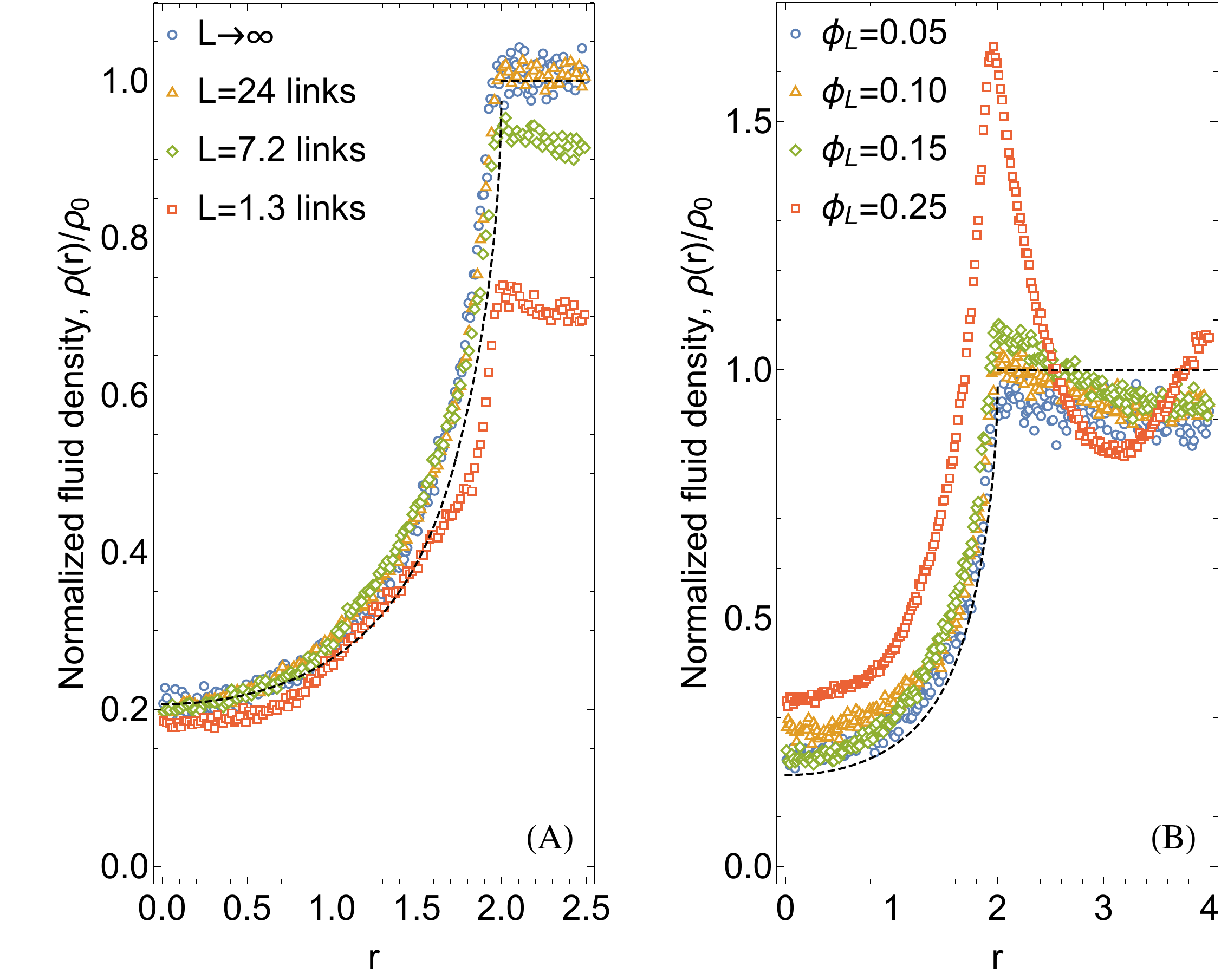}
    \caption{Comparison of theory (black dotted lines) that neglects bending and fluid-fluid interactions, with simulations (dots) that does not neglect bending (A), and include fluid-fluid interactions (B). We set (\(\phi=0.5\)) and vary persistence length $L$ and solvent density \(\phi_L = \pi \sigma^2 \rho/4\), where \(\sigma\) is the solvent diameter.}
    \label{Fig:Comparison}
\end{figure}

Another view, which we use to find the entropic force, is that \(p_\phi(r) \equiv p_\phi(r, \theta=0)\) is the expected free volume \(r\) away from the chain, see Fig. \ref{Fig:Polymer} A. Indeed, our simulations confirm that even for moderately flexible polymers,'' \(p_\phi(r)\) is a good approximation of the pair correlation between the polymer and the liquid (Fig. \ref{Fig:Comparison}). \added{Figs. \ref{Fig:Polymer}A and \ref{Fig:Comparison}A, B, were generated by molecular dynamics simulations of flexible or inflexible polymers immersed in a fluid. See \ref{Appendix:Simulation} for more details on how the simulations were performed.}

Note that the fluid density profile predicted by (\ref{ApproachProbability}), \(p_\phi(r)\), is different than that of an ordered chain, where the monomers are evenly spaced. For an ordered chain, the approach probability is \(p_\phi^{(\text{u})}(r) = \Xi(1-\phi\, \gamma(r))\),
which is significantly different for \(r<\sqrt{3}\). Interestingly, we see that the approach probability for an ordered chain is always less than or equal to that of a disordered chain with the same average density (Fig. \ref{Fig:Polymer}B).

Of course, the interpretation of (\ref{ApproachProbability}) as free volume or pair correlation function only holds exactly when the liquid density \(\phi_L = N v/V\) is small. When the liquid density is large, packing effects and sphere-sphere exclusion can come into play, and the density profile of the liquid will deviate from \(p_\phi(r)\) (Fig. \ref{Fig:Comparison}B).

\textbf{Depletion force between \added{rigid} disordered chains.} Now that we know the free volume near a disordered polymer, we can calculate the depletion force between two \added{rigid} polymer chains. 
While we present the derivation for polymers in 2D, the idea is the same for 3D (see \ref{Appendix:ThreeDimensions}).
Consider two parallel disordered lines of length \(L\) in two dimensions with sphere densities \(\phi\) and \(\varphi\) a distance \(X = (2 + x)R\) away from one another. Suppose that the lines are in a hard sphere fluid, also with radius \(R\), and with number density \(\rho = N_s / V\), total system volume (area) \(V\), and temperature \(T\). If the lines are closer than \(x=2\), the excluded volumes of the spheres on separate lines can overlap, resulting in an entropic force. The entropic force will depend the arrangement of spheres on each line, which are random variables, but we can calculate the average entropic force between the lines (which becomes exact as the line length \(L \to \infty\)) using (\ref{ApproachProbability}). From here on, by \(p_\phi(x)\), we mean \(p_\phi(x, 0, 0)\), i.e. \(\theta=0, \kappa=2\).

If the lines themselves cannot interact (\(x>0\)), then the arrangements of spheres on each line is independent. The expression \(p_\phi(x)\), (\ref{ApproachProbability}), tells us the probability that a point a distance \(x\) from a line can be occupied by a sphere. Letting \(r R\) denote the distance of a point from the left line, \(r^\prime R = (2+x-r)R\) is the distance of that point from the right line. For points at distances \(0<r<x\), a sphere can only be excluded by spheres on the left line, the probability that a sphere can occupy the volume is \(p_\phi(r)\). For points at distances \(x<r<2\), a sphere can be excluded by spheres on either line, so the probability that a sphere can occupy the volume is \(p_\phi(r) p_\varphi(x+2-r)\). Finally, for points at distances \(2<r<2+x\), a sphere can only be excluded by the right line, so the probability that a sphere can occupy the volume is \(p_\varphi (x+2-r)\).

\textbf{Depletion Force.} Each of these expressions involving \(p\)'s is the expected free volume per length at points between the lines. From this, we can calculate the free energy of the system, and then the entropic force. The partition function is \(Z(x) = V_E(x)^N/(N! \Lambda^{2N})\), where \(\Lambda = h/\sqrt{2\pi m k_B T}\) is the kinetic part (the \(h\) in \(\Lambda\) is Planck's constant), and \(V_E(x)\) is the free volume of the system. The expected free volume of the system is
\begin{align}
    V_E(x) = V - v_{\text{out}} - L x + L \lambda(x).
\end{align}
The term \(v_{out}\) corresponds to the reduced expected volume to the left of the left chain and right of the right chain, which can also be expressed in terms of \(p\), but does not depend on \(x\), so it will not matter to the entropic force calculation. In the third and fourth term, we subtract the volume between the lines, and add back the expected volume between the lines. From the considerations in the last section, we know that
\begin{align*}
    \lambda(x) &\equiv \int_0^{x+2} p_\phi(r) p_\varphi(r^\prime) dr 
    \\
    &= \int_0^x p_\phi(r)  dr + \int_x^2 p_\phi(r) p_\varphi(r^\prime)  dr + \int_2^{2+x} p_\varphi(r^\prime)  dr
\end{align*}
where \(r^\prime \equiv x+2-r\). The free energy \(F = -k_B T \log Z\) is
\begin{align}
    F(x) = F_0 - N k_B T \log\left[1 - \frac{v_{\text{out}}}{V} - \frac{L}{V} \left(x-\lambda(x) \right) \right]\nonumber
\end{align}
where \(F_0\) is independent of \(x\).
In the thermodynamic limit \(V\to\infty\), we use \(\log(1+\epsilon) \simeq \epsilon\) to get
\begin{align*}
    F(x) \simeq F_0^\prime - \rho  k_B T \left[x-\lambda(x) \right] L.
\end{align*}

From this, we find that the force per unit length \(\mathcal{P}(x)=-L^{-1} \partial F/\partial x\), and disorder averaged potential of the force \(u(x)\) are
\begin{align}
    \mathcal{P}(x) &= \rho  k_B T \left[ 1 - p_\varphi(x) + \int_x^2 p_\phi(r) \partial_r p_\varphi (r^\prime)  dr \right] \label{EntropicPressure}
    \\
    u(x) &= -\rho  k_B T [x-\lambda(x)]L
    \label{PMF}
\end{align}
which is our second main result. Note that the expression is the same if \(\phi\) and \(\varphi\) are exchanged (and integrating by parts), and so \(\mathcal{P}\) can be written in a symmetric form if that is preferred. Furthermore, \(\mathcal{P}\) is always negative, so the entropic force is attractive, and \(\mathcal{P}\) is zero for \(x>2\).

The correlation function for two chains can be easily found from (\ref{PMF}),
\begin{align}
    g_{ll}(x) = \exp \left( -\rho  k_B T \left[x-2-\lambda(x)+\lambda(2)\right] L\right).
\end{align}

\begin{figure}
    \centering
    \includegraphics[width=0.5\textwidth]{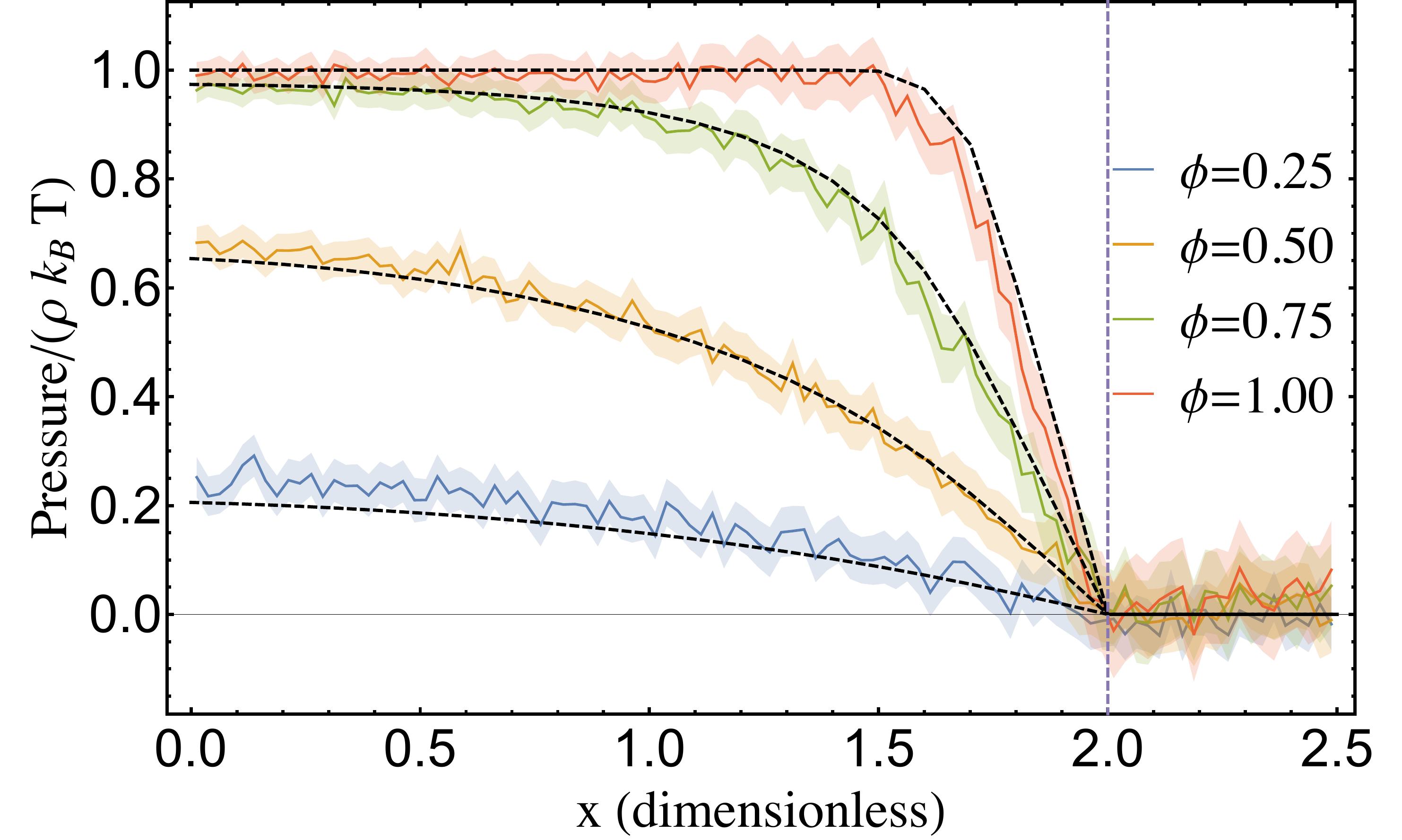}
    \caption{Comparing analytical formulas (dashed lines) to molecular dynamics simulations (solid lines). The standard deviation of the mean, \(\sigma_\mu = \sigma / \sqrt{N}\), of the simulations are marked with shaded regions representing \(2.5\,\sigma_\mu\). Pressures for systems with lines with density various densities, binned by line separation, \(x\). The red vertical dashed line marks \(x=2\), beyond which the entropic force vanishes.}
    \label{Fig:EntropicPressure}
\end{figure}

To verify our result, \added{(\ref{EntropicPressure}),} we ran molecular dynamics simulations \footnote{Using the GFlow molecular dynamics package, \url{https://github.com/nrupprecht/GFlow}} for \(\phi=\varphi\) measuring the force on disordered lines, and binning this by line separation. The lines were arranged to be rigid, and constrained to move only in the horizontal direction in order to measure the attractive force accurately (Fig. \ref{Fig:EntropicPressure}).

\added{As molecular dynamics requires forces, we simulate the hard spheres as ``very stiff elastic spheres'', which turns out to be very good approximation.} See \ref{Appendix:Simulation} for more simulation details.

\textbf{\added{Depletion force between flexible disordered chains}.} \added{We evaluate the entropic force between flexible chains using molecular dynamics simulations (Fig.\ref{Fig:EntropicPressure}A,B). Since we construct flexible chains of macromolecules instead of rigid linear configurations of macromolecules, the horizontal distance between chains is now a stochastic quantity both in position and time.
} 

\added{To get around this problem, we calculate the force on a per particle, i.e. the depletion pressure, which is the time averaged pressure \(F/(2R)\) on a particle as a function of distance, \(x\), from the closest point of the other chain. This way, we can also meaningfully compare depletion pressure of a flexible polymer with the net force on the rigid polymer per $2R$ length %. For the case of rigid linear polymers, \(\hat{n} = \pm\hat{x}\) (for the left and right polymers), and the average directed force per \(2R\) measured in the way we just described results in the same average pressure shown in Fig. 
 (cf. Fig.\ref{Fig:EntropicPressure}).}

\begin{figure}
    \centering
    \includegraphics[width=0.5\textwidth]{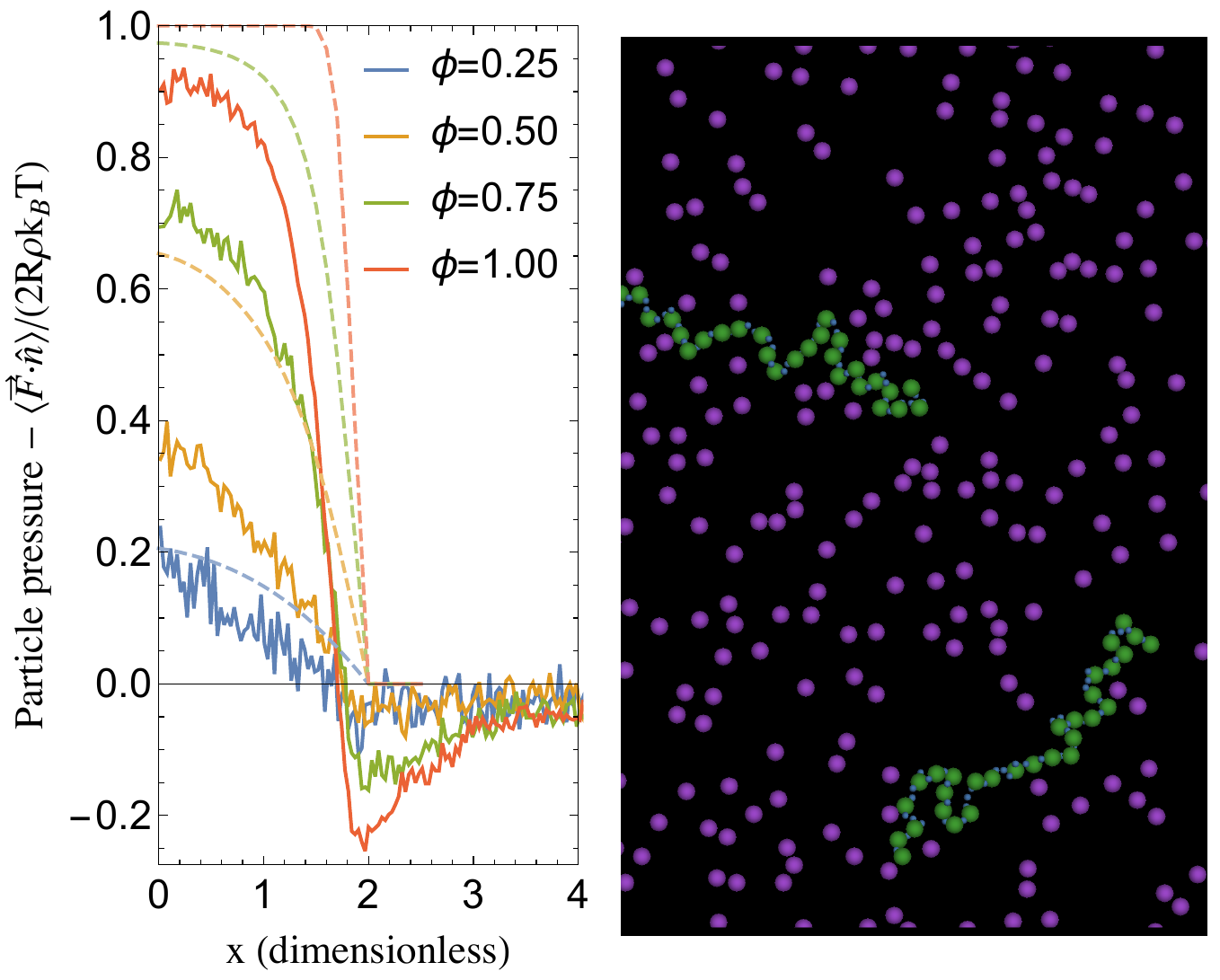}
    \caption{\added{(A) Entropic pressure between flexible chains. The dotted lines are the pressure predictions for a rigid linear polymer, (\ref{EntropicPressure}). While this is clearly not an accurate predictor of the magnitude of the pressure for the flexible chain, the general shape of the pressure curves are the same for \(x<2\). (B) Snapshot taken from a simulation of two flexible polymers. }}
    \label{Fig:FlexiblePressure}
\end{figure}

\added{While the entropic pressure between flexible polymers is similar to the entropic pressure between rigid linear polymers, several interesting differences are apparent (Fig. \ref{Fig:FlexiblePressure}). One is that the force between flexible chains are always less than that between the rigid chains. A second interesting feature is that while the force between rigid chains is zero beyond \(x=2\), for flexible chains it becomes repulsive.
}

\added{The repulsive part of the force is due to the fact that a flexible chain has many configurations it can be in, unlike a rigid chain. A snapshot of the instantaneous configuration of the two chains can be seen in Fig. \ref{Fig:FlexiblePressure} B. When two chains are near each other, they block each other from folding in as many ways as they could if they were far away from one another. This leads to an entropic repulsion term, which is what causes the pressure to be smaller overall than that between rigid polymers. Since the entropic repulsion depends on the number of allowable configurations for the entire chain, an entropic force can be felt by particles that are not close enough to feel the \emph{fluid} entropic force that is due to fluid arrangements (i.e. particles with \(x>2\)). The force is entirely due to the \emph{chain} entropic force.
}

\textbf{Connection to RISM.} Our method can be seen as an alternate route to solving the Reference Interaction Site Model (RISM) equation, a technique for finding correlation functions and depletion forces that is common in the polymer physics literature. The inhomogeneous RISM equation \cite{ishizuka2008integral} reads
\begin{align}
h (\mathbf{r}_{01}) = \int d \mathbf{r}_2 d\mathbf{r}_3 \omega(\mathbf{r}_{02}) C(\mathbf{r}_{23}) \left[ \omega(\mathbf{r}_{31}) + \rho h(\mathbf{r}_{31}) \right]
    \label{RISM}
\end{align}
where \(h\) is the matrix of total correlation function, \(\omega\) is the matrix of intramolecular correlation functions, and \(c\) is the matrix of direct correlation function. We use the inhomogeneous version of RISM since the fixed vertical lines breaks radial symmetry. The pair correlation function is \(g(\mathbf{r}) = h(\mathbf{r})+1\).

In our system, the correlation functions are random variables that depend on the arrangement of both chains. Suppose the chains, \(l\), \(l^\prime\), are parametrized so their sites have positions \(\{p^{(m)}_x, p^{(m)}_y\}\), \(m\in \{l,l^\prime\}\). Since we want to find the correlation between just the \(x\) coordinates of the lines, we will use the molecular correlation between \(\{p^{(m)}_x, 0\}\) and the sites in the line as \(\omega_{m,i}\), as opposed to the typical \(\omega_{m,ij}\) correlation between sites \(i\) and \(j\). For two random chain configurations \(\alpha = \{h^{(l)}_1, \dots h^{(l)}_N; h^{(l^\prime)}_1, \dots h^{(l^\prime)}_N \}\) where \(l, l^\prime\) index the two chains, this is 
\begin{align*}
    \omega_{m,k}^{(\alpha)}(\mathbf{r}) = \delta(r_y - h^{(m)}_k + p^{(m)}_y) \times \delta(r_x- p^{(m)}_x).
\end{align*}

In our model, where the chains do not directly interact in our range of interest of \(h\), \(C_{ll^\prime,ij}(\mathbf{r}) = 0\). Inserting these functions in (\ref{RISM}) and keeping only terms of order \(\rho\), the total correlation function between \(p_x^{(l)}, p_x^{(l^\prime)}\) is
\begin{align*}
    &h_{ll^\prime}^{(\alpha)}(\Delta p_x) \!= \!\rho\! \sum_{j,k=1}^{N} \int_{\mathbb{R}^2} \!\!\!d\mathbf{r}_s C^{(\alpha)}_{ls,j}\left((x_s - p^{(l)}_x) \hat{x} + (y_s - h^{(l)}_{j}) \hat{y} \right) 
    \\
    & \times C^{(\alpha)}_{sl^\prime,k}\left((p^{(l^\prime)}_x - x_s) \hat{x} + (h^{(l^\prime)}_{k} - y_s) \hat{y} \right).
\end{align*}

Hereafter, we abbreviate the vector positions in the \(C\) functions as \(\mathbf{r}_{l,j}\), \(\mathbf{r}_{m,k}\). Note that \(C_{ls}^\alpha\), \(C_{sl^\prime}^\alpha\) are \(N\) component vectors, one entry for each interaction site on the corresponding line.

Often, this is where the analytical part of the RISM procedure stops, the equations for \(h\) and \(C\) are evaluated using Picard iteration or some similar technique, leading to the radial correlation function. But our object of study is \(u(r) \equiv \langle U^{(\alpha)}(r) \rangle_\alpha\), not the radial correlation function or potential of mean force for any \emph{specific} random polymer arrangement. However, the potential of mean force for long chains should converge to the disorder averaged potential of mean force in the thermodynamic limit.

By definition, \(-\beta U^{(\alpha)}(r) = \log \left(1+h^{(\alpha)}(r)\right)\). Recalling that \(\rho\) is small, and averaging over the disorder,
\begin{align*}
    & -\beta u(x)= \left\langle \log \left(1+h^{(\alpha)}(x) \right) \right\rangle = \langle h^{(\alpha)}(x) \rangle + \mathcal{O}(\rho^2) 
    \\
    &= \rho \sum_{j,k=1}^N \int_{\mathbb{R}^2} d\mathbf{r}_s \left\langle C_{ls,j}^{(\alpha)}(\mathbf{r}_{l,j}) C^{(\alpha)}_{sl^\prime,k}(\mathbf{r}_{m,k}) \right\rangle
    \\
    &= \rho \int_0^{h+2} dx_s \int_0^L dy_s \left\langle \sum_{j,k} C_{ls,j}^{(\alpha)}(\mathbf{r}_{l,j}) C^{(\alpha)}_{sl^\prime,k}(\mathbf{r}_{m,k}) \right\rangle.
\end{align*}
Note that the smallness of \(h\) is an essential element in the simplification of this problem, since for most probability distributions, \(\langle \log f(x,\omega) \rangle \ne \log \langle f(x,\omega) \rangle\).

Concentrating on the inner integral, which ignores edge effects, and noting that the two correlation functions depend on different and independent chain arrangements,
\begin{align*}
    \chi(x) &\equiv \int_0^L dy_s \left\langle \sum_{j} C_{ls,j}^{(\alpha)}(\mathbf{r}_{l,j}) \right\rangle \left\langle \sum_{k} C^{(\alpha)}_{sl^\prime,k}(\mathbf{r}_{m,k}) \right\rangle 
    \\
    &= L c_{ls}(x-p^{(l)}_x)  c_{ls}(p^{(l^\prime)}_x - x)
    \\
    c_{ls}(x-p^{(l)}_x) &\equiv \left\langle \sum_{j} C_{ls,j}^{(\alpha)}(\mathbf{r}_{l,k}) \right\rangle 
\end{align*}
Going from the first to second line is a consequence of the fact that if \(L\) is large, or when we are using periodic boundary conditions, the \(C\)s will on average be independent of \(y\), and the disorder of each chain can be averaged over independently, so the disorder averaged sum of \(C\)s is not a function of \(y\).

We now have to determine the vector or direct correlation functions, \(C^{(\alpha)}_{ls}(x)\). For hard spheres at low density, the direct correlation function is known to be \(\Theta(2R-r)\) \cite{tejero2007direct}, which makes physical sense given the nature of the hard sphere interaction. Due to the fact that our disordered chain sites' can be close to each other and strongly correlated, the correlation function will differ from this. Physically speaking, \(\sum_i C_i(\mathbf{r})\) should sum to \(1\) in regions of space into which the hard sphere fluid cannot penetrate, and 0 where it can. Therefore, a physically realistic choice when the liquid is dilute is
\begin{align*}
    C^{(\alpha)}_{ls,i}(\mathbf{r}) = 
    \begin{cases}
    0 & \| \mathbf{r} - \mathbf{p}_i\| \ge 2 R \\
    -\Big[\sum_j \Theta \left( 2 R - \|\mathbf{r} - \mathbf{p}_j \| \right) \Big]^{-1} & \| \mathbf{r} - \mathbf{p}_i\| < 2 R 
    \end{cases}.
\end{align*}
This evenly splits the responsibility of excluding liquid particles among all sites that exclude that region of space.

Since \(\langle \sum_j C^\alpha_{ns}(x) \rangle\) just counts the probability that a random configuration excludes a hard sphere from being at a distance \(x\) from the chain, we get 
\begin{align*}
    \chi(x) &= L [1-p_\phi(x)] [1-p_\varphi(x)]
    \\
    -\beta  u(x) &= \rho  L  \int_0^{x+2} dr   [1-p_\phi(r)] [1-p_\varphi(r)].
\end{align*}
Expanding the terms in the integral, we get that
\begin{align*}
    &\frac{u(x)}{\rho k_B T L} = -x + \int_0^{x+2} p_\phi(r) p_\varphi(x+2-r)  dr 
    \\
    &- \Big( 2 - \int_0^2 p_\phi(r) dr - \int_0^2 p_\varphi(r)   dr \Big).
\end{align*}
Note that the term in parenthesis is a constant (for fixed \(\phi, \varphi\)), so it does not matter in the averaged potential of mean force and we will omit it. Some simplification arise due to the fact that \(p_{\phi,\varphi}(r) = 1\) for \(2\le r\). We have, therefore, rederived our expression via RISM, which matches our expression for free energy (up to constants that do not depend on \(h\)), \(u(x) = -\rho k_B T [x-\lambda(x)] L\).

\textbf{Conclusion.} Our two main \added{theoretical} results are the fluid density profile surrounding a disordered \added{straight} polymer, and depletion force between two such chains (eqn. (\ref{ApproachProbability}, \ref{EntropicPressure}))\added{. Additionally, we have numerically determined the depletion force between flexible polymers}. While there are many studies on the interactions between polymers with definite structure functions, here we focused on \emph{disordered} polymers, with random structure functions. We evaluated the average correlation between a disordered chain and a hard sphere fluid (i.e. the approach probability, \(p_\phi\)), and used this to find the depletion force between disordered chains in an approach similar to that of Asakura-Oosawa equations. We also showed how RISM can be extended, in the low density limit, to derive the average mean potential for this random ensemble.

It would be interesting to extend the random RISM method presented here to take into account fluids in the high density limit. There has been success in spin glass theory in dealing with averages over quenched disorder, e.g. the replica approach, which is also used in replica density functional theory \cite{reich2004replica,schmidt2005replica} and replica Ornstein-Zernike theory \cite{pizio1997adsorption} to study fluids in random porous materials. A similar approach might be fruitful here.

\clearpage

%---> Appendices
\onecolumngrid

% Label by letter. Restart from 'A'
\setcounter{subsection}{0}
\renewcommand{\thesubsection}{Appendix \Alph{subsection}}

\subsection{Calculation details}
\label{Appendix:CalcDetails}

To evaluate (\ref{PassageProbabilityIntegralForm}) when \(\xi(r) > 1\) we exchange the order of the sum and integrals, and look at each term in the sum,
\begin{align}
p_\phi(r, N) &= \frac{1}{L \Gamma_N} \sum_{j=1}^N T_j(r)
\label{PInTermsOfT}
\\
T_j(r) &\equiv \int_{x_0+2R}^{L-2R(N-1)} \hspace{-0.2in}dx_1 \dots \int_{x_{j-1}+2R}^{L-2R(N-j)} \hspace{-0.2in}dx_j   \Xi \big( x_j - x_{j-1} - 2R \xi(r) \big) \int_{x_j+2R}^{L-2R(N-j-1)} \hspace{-0.2in}dx_{j+1} \dots \int_{x_{N-2}+2R}^{L-2R} \hspace{-0.2in}dx_{N-1}.
\label{TFunction}
\end{align}
But since our choice of which sphere was 0 is arbitrary, and we are using periodic boundary conditions (so there is circular symmetry), we must have that \(T_j(r) = T_k(r)\) for all pairs \(j, k\). In other words, all the \(T_j\) are equal, call this common value \(T(r)\). We will evaluate the one that requires the least work, \(T(r) = T_1(r)\), since we can integrate all the integrals after the \(\Xi\) function inductively.
\begin{align}
T(r) &= \int_{x_0+2R}^{L-2R(N-1)} dx_1 \Xi \big( x_1 - x_0 - 2R \xi(r) \big) \int_{x_1+2R}^{L-2R(N-2)} dx_2 \dots \int_{x_{N-2}+2R}^{L-2R} dx_{N-1}
\nonumber 
\\
&= \int_{x_0+2R}^{L-2R(N-1)} dx_1  \frac{\Xi \big( x_1 - x_0 - 2R \xi(r) \big)}{(N-2)!} \times (L - 2R(N-1) - x_1)^{N-2}
\end{align}
We can deal with the \(\Xi\) function by adjusting the bounds of integration. Making the necessary adjustments,
\begin{align}
T(r) &= \frac{1}{(N-2)!} \int_{x_0+2R \xi(r)}^{L-2R(N-1)}dx_1 \left(x_1 - x_0 - 2R \xi(r) \right) \left(L - 2R(N-1) - x_1\right)^{N-2} 
\label{AdjustBounds}
\end{align}
if \(\xi(r) - 1 < N (1/\phi - 1)\), and is \(0\) otherwise.

It is safe to exchange the lower bound of integration in (\ref{TFunction}), \(x_0 + 2R\), with \(x_0 + 2R \xi(y)\) since we are treating the case that \(\xi(r)>1\). The second case comes from the fact that \(\xi(r) - 1 \ge N(1/\phi -1) \) means that the lower bound of integration would have to be above the upper bound. The \(T\) integral can be performed as follows, call \(\alpha \equiv 2R \xi(r)\) and \(\beta \equiv L-2R(N-1)\). Then
\begin{align}
T(r) = \int_\alpha^\beta dx  (x-\alpha) (\beta - x)^{N-2}
= \int_0^{\beta-\alpha} du   (\beta-\alpha-u) u^{N-2}
= \frac{(\beta-\alpha)^{N-1}}{N(N-1)}
= \frac{L^N}{N(N-1)} \left( 1 - \phi \left[1 + \frac{\xi(r)-1}{N}\right]\right)^N
\end{align}
where we made the substitution \(u = \beta-x\). The integral is then just two polynomial integrals which can be performed with ease.
\begin{align}
T(r) = \frac{L^N}{N!} \left( 1 - \phi \left[1 + \frac{ \xi(r)-1 }{N} \right] \right)^N 
\end{align}
if \(\xi(r) - 1 < N (1/\phi - 1)\), and is zero otherwise.
This can be put into (\ref{PInTermsOfT}) and gives us our approach probability for \(\xi(r) > 1\), and can be combined with (\ref{ApproachProbabilityFar}) to yield our complete expression for approach probability.

\subsection{Calculation without periodic boundary conditions}
\label{Appendix:DifferentBCs}

Here, we give some details about the approach probability for a chain, without periodic boundary conditions. In the thermodynamic limit, the probability is (\ref{ApproachProbability}), but for finite \(N, L\), there are corrections of order \(\mathcal{O}(1/N)\), and the problem is more complicated. For simplicity, we assume that the spheres are of equal size (so \(\kappa=2\)), and that \(\theta=0\).

Let \(\Gamma^{(l)}_{N,R,L}\) be the measure of configurations of \(N\) spheres of radius \(R\) on a line of length \(L\) such that no sphere extends ``beyond'' the line - that is the centers of the spheres must fall in the range \([R, L-R]\). We suppose that the line is along the x axis, from 0 to \(L\). Using the same methods used to solve the analogous problem with periodic boundary conditions, we get \(\Gamma^{(l)} = \frac{L^N}{N!}(1-\phi)^N\).

A sphere launched towards the chain can collide if it has x coordinate in the range \([-R, L+R]\), so we consider our launched spheres to be chosen uniformly at random from this range. A sphere with x coordinate less than \(x_1\) can only interact with the first sphere, likewise, a sphere with x coordiate greater than \(x_N\) can only interact with the last sphere. 
\begin{align*}
    p_\phi^{(l)}(r, N) &= \frac{1}{(L+2R) \Gamma^{(l)}_N} \left( L(r) + \sum_{j=2}^N S_j(r) + R(r) \right)
    \\
S_j(r) &\equiv \int \mathsf{D}x \Xi \left(x_j - x_{j-1} - 2 R \gamma(r) \right), \qquad    
    L(r) \equiv \int \mathsf{D}x \left(x_1 + 2 R - R \gamma(r) \right),\\    
    R(r) &\equiv \int \mathsf{D}x \left( L - x_N - R \gamma(r) \right), \qquad
    \int \mathsf{D}x \equiv \prod_{k=1}^N \int_{x_{k-1}}^{L-2(N-k+1)R}  dx_k   \vert_{x_0 = -2R}.
\end{align*}
In the case where \(\gamma < 1\), the \(L(r)\), \(R(r)\) integrals can be evaluated (they have the same value, by symmetry)
\begin{align*}
    L(r) = R(r) = \frac{L^{N+1}}{(N+1)!} (1-\phi)^N \left[ 1 + (1-\gamma/2) \frac{\phi}{N} - \frac{1}{2} \phi \gamma \right] \equiv E(r)
\end{align*}
and the sums telescopes and so can be evaluated,
\begin{align*}
    p^{(l)} (r) = \frac{1}{1+\frac{\phi}{N}} \left( 1- \phi \gamma(r) - \frac{\phi}{N} \right)
\end{align*}

For the \(\gamma>1\) case, all the \(S_j\) are equal, like before, and \(L(r)=R(r)\) by symmetry. By adjusting the integration bounds in the integral, just as in (\ref{AdjustBounds}), we can evaluate \(S(r)\). 
\begin{align*}
    S(r) &= \frac{L^{N+1}}{(N+1)!} (1-\phi)^{N+1} \left( 1 - \frac{\phi}{1-\phi}(1-\gamma) \frac{1}{N} \right)^N \left\{ 1 - \frac{\phi}{1-\phi}(\gamma-1) \frac{1}{N} \left[ 1 - \left(
    \frac{\phi}{1-\phi} \frac{\gamma-1}{1-\frac{\phi}{1-\phi}(\gamma-1)\frac{1}{N}} \frac{1}{N}
    \right)^N \right] \right\}
\end{align*}

Putting all this together, we find that the approach probability to a chain without periodic boundary conditions is
\begin{align*}
    p^{(l)}(r) = \begin{cases}
    \frac{1}{1+\frac{\phi}{N}} \left( 1- \phi \gamma(r) - \frac{\phi}{N} \right) & \gamma(r) \le 1
    \\
    \frac{1}{(L+2R)\Gamma^{(l)}} \left( 2 E(r) + (N-1) S(r) \right) & \gamma(r) > 1
    \end{cases}.
\end{align*}
While this equation is much more complex than its periodic boundary condition counterpart, they both have the same limiting value in the thermodynamic limit.

\subsection{Free energy in three dimensions}
\label{Appendix:ThreeDimensions}

For a pair of disordered chains in three dimensions, the procedure for finding the free volume is similar to the case of disordered parallel chains. Suppose the fluid radii, and the radii of the monomers are both \(R\). Let \(V_A\) be the volume within \(2 R\) of the first chain, \(V_B\) be the volume within \(2 R\) of the second chain, and \(V_{\text{int}} \equiv V_A \cap V_B \). The expected free volume is now a complicated function of the positions of the centers of the chains, \(\mathbf{r}_1, \mathbf{r}_2\), and normal vectors describing the orientations of the chains, \(\hat{\mathbf{n}}_1\), \(\hat{\mathbf{n}}_2\).
\begin{align*}
    V_E(\{\mathbf{r}, \mathbf{n}\}) &= V - \left[V_A + V_B - V_\text{int} (\{\mathbf{r}, \hat{\mathbf{n}}\}) \right] + \int_{V_A} d\mathbf{s}  P_\phi(\mathbf{r}_1, \hat{\mathbf{n}}_1 ; \mathbf{s}) + \int_{V_\text{int}} d\mathbf{s}  P_\phi(\mathbf{r}_1, \hat{\mathbf{n}}_1 ; \mathbf{s}) \cdot P_\varphi(\mathbf{r}_2, \hat{\mathbf{n}}_2 ; \mathbf{s}) + \int_{V_B} d\mathbf{s}  P_\varphi(\mathbf{r}_2, \hat{\mathbf{n}}_2 ; \mathbf{s})
    \\
    P_\phi(\mathbf{r}, \hat{\mathbf{n}}; \mathbf{s}) &\equiv p_\phi \left( \| \mathbf{s} - \mathbf{r} - \left((\mathbf{s} - \mathbf{r})\cdot \hat{\mathbf{n}}\right) \hat{\mathbf{n}} \|/R \right).
\end{align*}
The function \(p_\phi\) is the familiar probability of approach from before. Since the partition function is \(Z(\{\mathbf{r}, \hat{\mathbf{n}} \}) = V_E(\{\mathbf{r}, \hat{\mathbf{n}} \})^N / (N!   \Lambda^N)\), and \(F = -k_B T \log Z\), we take the log and expand in terms of \(1/V\),
\begin{align*}
    F(\{\mathbf{r}, \hat{\mathbf{n}} \}) = F_0- \rho  k_B T \left( V_\text{int} (\{\mathbf{r}, \hat{\mathbf{n}}\})  + \int_{V_A} d\mathbf{s}  P_\phi(\mathbf{r}_1, \hat{\mathbf{n}}_1 ; \mathbf{s}) + \int_{V_\text{int}} d\mathbf{s}  P_\phi(\mathbf{r}_1, \hat{\mathbf{n}}_1 ; \mathbf{s}) \cdot P_\varphi(\mathbf{r}_2, \hat{\mathbf{n}}_2 ; \mathbf{s}) + \int_{V_B} d\mathbf{s}  P_\varphi(\mathbf{r}_2, \hat{\mathbf{n}}_2 ; \mathbf{s}) \right)
\end{align*}
where \(F_0\) is a term that does not depend on the configuration of the cylinders.

Unfortunately, after this point, further simplification, becomes impossible. To find the depletion force on the chains, we evaluate the gradient of the free energy,
\begin{align*}
    \mathcal{F}(\{\mathbf{r}, \hat{\mathbf{n}} \}) = -\rho k_B T\cdot \nabla_{\mathbf{r}} \left( V_\text{int} (\{\mathbf{r}, \hat{\mathbf{n}}\})  + \int_{V_A} d\mathbf{s}  P_\phi(\mathbf{r}_1, \hat{\mathbf{n}}_1 ; \mathbf{s}) + \int_{V_\text{int}} d\mathbf{s}  P_\phi(\mathbf{r}_1, \hat{\mathbf{n}}_1 ; \mathbf{s}) \cdot P_\varphi(\mathbf{r}_2, \hat{\mathbf{n}}_2 ; \mathbf{s}) + \int_{V_B} d\mathbf{s} P_\varphi(\mathbf{r}_2, \hat{\mathbf{n}}_2 ; \mathbf{s}) \right).
\end{align*}
The torque on the chains could be evaluated similarly. Note that the volumes over which we are integrating depend on the orientation of the particles. Note that \(F\) and the magnitude of \(\mathcal{F}\) only actually depend on the relative locations and orientations of the two chains.

As a final word of caution, this formula is only valid when the minimum distance between the lines is \(\ge 2 R\), which is necessary to ensure that the arrangement of spheres on the lines are independent from one another. Furthermore, edge effects are not included, so this formula assumes that the parts of the lines that are close to one another are away from the ends of the lines.

\subsection{Simulation details}
\label{Appendix:Simulation}

We give some further details concerning the molecular dynamics simulations used to obtain data for Fig.s \added{ \ref{Fig:Polymer}, \ref{Fig:Comparison}, \ref{Fig:EntropicPressure}. Straight polymers, the objects that we also treat theoretically in this paper, are generated by first specifying the length of the ``linear backbone'' of the polymer (which is not explicitly represented by objects in the simulation, but helps us keep track of where the monomers should go), \(L\), and a target linear volume fraction, \(\phi = 2 N R/L\). Only certain volume fractions are possible, the ones that correspond to integral numbers of disks on the backbone. The total number of spheres, \(N\), is calculated from \(\phi,\,R=0.05\).
}

\added{The positions of the disks on the backbone are chosen by picking \(N\) numbers uniform at random from \([0,L-2RN]\) and sorting them so \(x_1<...<x_N\). The positions of the disks are set to \(r_1 \equiv x_1 + R\), \(r_2 \equiv r_1 + x_2 - x_1 + 2 R\), ... , \(r_N = r_{N-1} + x_N - x_{N-1} + 2R\). This method samples uniformly at random from the set of all configurations of disks on the backbone with the constraint that no disks overlap.
}

\added{To create a flexible polymer, as we do to obtain data for Fig. \ref{Fig:Polymer} and \ref{Fig:FlexiblePressure}, we modify the above algorithm slightly. We pick a small monomer radius, \(R_c = 0.02\), for the disks that act as chain links in the polymer. The chain disks do not interact with any other particles except via harmonic bond and angle forces with their (at most) two adjacent neighbors in the chain. Again, based on the desired chain length and linear volume fraction, the total number of (large) disks is calculated. Positions for the large disks are calculated as before, but since we are filling the space between large disks with chain monomers, we calculate the expected number of chain particles between the two large disks, \(n_x = (r_k - r_{k-1})/(2R_c) + 1\) and place \(n_c = \text{floor}(n_x) + \hat{C}\) chain particles between adjacent large disks. The random variable \(\hat{C}\) is \(1\) with probability \((n_x - n_c) \in [0,1]\) and 0 otherwise. This compensates on average for the fact that we always round \(n_x\) down to get the deterministic part of \(n_c\).
}

\added{As mentioned before, the hard sphere interaction was approximated by using a harmonic repulsion force with a very large stiffness. This} force was used for chain-chain, chain-solvent, and solvent-solvent forces, 
\[
\mathbf{F}(r) = - \kappa \, \Theta(R_1 + R_2 - r) \mathbf{\hat{r}}.
\]
where \(R_1,\,R_2\) are the radii of the interacting spheres, \(\kappa\) is the repulsion constant, and  \(\Theta\) is the Heaviside step function. \added{In our simulations, the radii of all the interacting particles (the large monomers and the hard sphere fluid) are the same, \(R=0.05\), each particle had density \(1\) (and therefore mass \(m=0.00785\)), and the spring stiffness used was \(\kappa=500\). The period of collision was therefore about \(t_c = 0.025\). In all simulations, periodic boundary conditions were used, and time evolution is carried out using Velocity-Verlet integration, modified to run a Nose-Hoover thermostat \cite{evans1985nose} to keep the system at constant temperature. To evaluate non-bonded forces, a linked cells structure is used to create Verlet list for force evaluation.}

The \added{straight} polymers are each rigid disordered chains\added{, created as described above, which we constrain} to only move in the x direction, and remain vertical, all net torque and force in the \(y\) direction are set to zero.
The right chain was displaced a random amount in the \(y\) direction, uniformly chosen to be between 0 and \(2 R\) above the start of the left chain. This insures that measurements are not effected by systematic correlations between the spheres on the two chains, which can occur at high \(\phi\). At distances greater than \(5 R\), a harmonic force was activated to prevent the chains from departing far away thereby allowing us to collect larger amount of data. 

\added{For the straight polymer,} each \(\phi\) curve was composed using data from 18 simulations, 6 where the chain were in square boxes of solvent of side length \(160R\), with 408 solvent particles, and 12 where the chains were in boxes of solvent of side length \(320R\), with 1630 solvent particles. In both cases, the length of the lines were \(32R\), and the volume density of the solvent was \(\phi = \frac{\pi R^2}{V} = 0.05\).

\added{The pressure for the straight chains is calculated by binning the total force on each chain, projected in the \(\pm \hat{x}\) direction, towards the other chain. Each bin is averaged, and the force is divided by the length of the line, \(L\).}

\added{As alluded to in the main text, calculating the pressure for the flexible chains is more complicated since the orientation of each chain segment is different, and the chains are no longer parallel. 
For each monomer (disk) in each chain, we calculate the nearest point on the flexible backbone of the other chain, which may lie between two monomers. If this is the case, we assume that the chain is straight between monomers. We then bin the force on the monomer projected in this direction, \(\hat{n}\), divided by \(2R\) (this is the pressure in the \(\hat{n}\) direction on the particle), binning by the distance to that nearest point on the other chain.
%To see how this procedure compares to the straight polymer, if we assume that there is some average pressure exerted on the chain, then the force it causes on a single monomer of the chain is \(2 R \mathcal{P} = \langle \mathbf{F} \cdot \mathbf{\hat{n}}\rangle\) (since the ``length'' of a particle is \(2R\)), and therefore \(\mathcal{P} = \frac{\langle \mathbf{F} \cdot \mathbf{\hat{n}}\rangle}{2R}\).
}

\added{Since the flexible polymers are not constrained to be near one another, have many more internal degrees of freedom, and because we use smaller simulations than in the linear polymer case, we average the system much more that in the straight polymer case to obtain accurate data. Each \(\phi\) curve was composed of 150 simulations run for 30,000 seconds each. Each simulation took place in a box of solvent with side length \(100R\) with 319 solvent particles. The length of the chain was \(40R\), and the density of the solvent was \(\phi = 0.1\).}

The GFlow molecular dynamics package is freely available at \url{https://github.com/nrupprecht/GFlow}.

\subsection{\added{Numerical approximation of entropic pressure}}
We have use \cite{eureqa} to find an approximate closed form expression for (\ref{EntropicPressure}). For $\phi=\varphi$,
\begin{align*}
\mathcal{P}(x)\!=\!(c_1 \!+\! c_2 x^4) \phi^2 \!\!-\! \tanh[(c_3 \!+\! x^2) \phi^2] (c_4 \!+\! c_5 \tanh[4 \phi^4])
\end{align*}
in units of $\rho k_BT$. With $c_1=-0.158, c_2=0.031, c_3=3.667, c_4=0.899$ and $c_5=0.236$ the mean error is $0.5\%$.

\bibliography{bibliography.bib}

\end{document}